\documentclass[twocolumn,aps,floatfix]{revtex4}
\usepackage{amssymb}
\usepackage{graphicx}

\begin{document}
\title{Soliton trains in Bose-Fermi mixtures}

\author{T. Karpiuk,$\,^1$ M. Brewczyk,$\,^1$
        S. Ospelkaus-Schwarzer,$\,^2$ K. Bongs,$\,^2$
        M. Gajda,$\,^3$ and K. Rz{\c a}\.zewski$\,^4$}

\affiliation{\mbox{$^1$ Uniwersytet w Bia{\l}ymstoku, ulica Lipowa 41, 15-424 Bia{\l}ystok, Poland}  \\
\mbox{$^2$ Institut f\"ur Laser-Physik, Universit\"at Hamburg,
Luruper Chaussee 149, 22761 Hamburg, Germany}  \\
\mbox{$^3$ Instytut Fizyki PAN, Aleja Lotnik\'ow 32/46, 02-668 Warsaw, Poland} \\
\mbox{$^4$ Centrum Fizyki Teoretycznej PAN, Aleja Lotnik\'ow 32/46, 02-668 Warsaw, Poland}  }

\date{\today}

\begin{abstract}
We theoretically consider the formation of bright solitons in a
mixture of Bose and Fermi degenerate gases. While we assume the
forces between atoms in a pure Bose component to be effectively
repulsive, their character can be changed from repulsive to
attractive in the presence of fermions provided the Bose and Fermi
gases attract each other strongly enough. In such a regime the
Bose component becomes a gas of effectively attractive atoms.
Hence, generating bright solitons in the bosonic gas is possible.
Indeed, after a sudden increase of the strength of attraction
between bosons and fermions (realized by using a Feshbach
resonance technique or by firm radial squeezing of both samples)
soliton trains appear in the Bose-Fermi mixture.
\end{abstract}

\maketitle

Solitonic solutions are a very general feature of nonlinear wave
equations. Solitons have been studied in many different physical
systems ranging from particle physics to optics. They differ from
ordinary wave packets as they retain their shape while propagating
instead of spreading due to dispersion. This intriguing feature is
based on the existence of a nonlinear interaction which
compensates for dispersion and produces a self-focusing effect on
the propagating wave packet.

Dilute atomic quantum gases offer a unique environment to study
fundamental solitonic excitations in a pure quantum system with
intrinsic nonlinearity. Since the interparticle interaction
causing this nonlinearity can be both attractive and repulsive,
the Gross-Pitaevskii equation describing the evolution of the
condensate wave function exhibits both dark and bright solitonic
solutions \cite{Zakharov}. Dark solitons as a fundamental
excitation in stable Bose-Einstein condensates with repulsive
interparticle interaction have been studied in different
geometries \cite{Bur1999,Den2000,And2001}.

Bright solitons have been observed in Bose-Einstein condensates of
$^{7}\mathrm{{Li}}$ in quasi-one-dimensional geometry
\cite{Khay2002,Hul2002}. However, in three-dimensional geometry
usually used to prepare the sample the necessary large and
negative scattering length leads to density-limited particle
numbers (dynamical instability -- collapse). The observation of
bright solitons was therefore only possible due to magnetic tuning
of the interactions from repulsive (used to form a stable
Bose-Einstein condensate) to attractive during the experiments.

Another experimental approach to bright matter wave solitons was
realized in the recently reported observation of gap solitons
\cite{Ob04} in a condensate with repulsive interactions by
engineering of the matter wave dispersion relation via
sophisticated manipulation in a periodic potential (concept of
negative effective mass \cite{Len1994}).

In this Letter we propose a novel scheme to realize bright
solitons in one-dimensional atomic quantum gases. In particular we 
study the formation of bright solitons in a Bose-Einstein condensate
embedded in a quantum degenerate Fermi gas. One important feature
is that this mixture allows tuning of the one-dimensional interactions 
not only by Feshbach resonances but also by simply changing the trap
geometry.

We consider the bare interaction between bosonic atoms to be
repulsive ($g_{B}>0$) whereas the particle interaction between
bosonic and fermionic atoms is assumed to be strongly attractive
($g_{BF}<0$). Bright solitons in Bose-Fermi gas mixtures are then
produced as a result of a competition between two interparticle
interactions: boson-boson repulsion versus boson-fermion
attraction. Experimentally, this situation is for example
accessible in Bose-Fermi mixtures of bosonic $^{87}\mathrm{{Rb}}$
atoms and fermionic $^{40}\mathrm{{K}}$ atoms.

We determine the critical strength of attraction between bosons
and fermions necessary for the formation of bright solitons and
show that these parameter regimes might be achievable in present
experiments. We study the formation of bright solitons by
different excitation mechanisms: either increasing the attractive
boson-fermion interaction by Feshbach resonance techniques or by
radial squeezing of the mixture which corresponds to increasing
the effective one-dimensional scattering length. We contrast the
response of the system following adiabatic and fast increase of
the boson-fermion interaction strength.

We consider a Bose-Fermi mixture confined in a trap at zero
temperature and describe this system in terms of the many-body
wave function $\Psi ({\bf x}_1,...,{\bf x}_{N_B};{\bf
y}_1,...,{\bf y}_{N_F})$,
where $N_B$ and $N_F$ are the numbers of bosons and fermions,
respectively. However, instead of direct (and of course
approximate) solving of the many-body Schr\"odinger equation we
start with the equivalent approach based on the Lagrangian
density. Since the fermionic sample is spin-polarized only
boson-boson and boson-fermion interactions are included. At zero
temperature we assume that the wave function of Bose-Fermi mixture
is a product of the Hartree ansatz for bosons and the Slater
determinant (antisymmetric wave function) for fermions

\begin{eqnarray}
&&\Psi ({\bf x}_1,...,{\bf x}_{N_B};{\bf y}_1,...,{\bf y}_{N_F}) =
\prod_{i=1}^{N_B} \varphi^{(B)}({\bf x}_i)
\nonumber \\
&&\times \frac{1}{\sqrt{N_F!}} \left |
\begin{array}{lllll}
\varphi_1^{(F)}({\bf y}_1) & . & . & . & \varphi_1^{(F)}({\bf y}_{N_F}) \\
\phantom{aa}. &  &  &  & \phantom{aa}. \\
\phantom{aa}. &  &  &  & \phantom{aa}. \\
\phantom{aa}. &  &  &  & \phantom{aa}. \\
\varphi_{N_F}^{(F)}({\bf y}_1) & . & . & . & \varphi_{N_F}^{(F)}({\bf y}_{N_F})
\end{array}
\right |    
\;.
\label{ansatz}
\end{eqnarray}

Now, all the single-particle wave functions $(\varphi^{(B)}$,
$\varphi_1^{(F)},...,\varphi_{N_F}^{(F)})$ have to be determined.
Therefore the many-body wave function (\ref{ansatz}) is inserted into
the Lagrangian density and integrated over the spatial coordinates to
get the Lagrangian. In fact, this integration is performed only over
$N_B-1$ bosonic and $N_F-1$ fermionic coordinates and leads in this
way to the mean-field single-particle Lagrangian. The Euler-Lagrange
equations corresponding to this Lagrangian are the basic equations of
the presented approach
\begin{eqnarray}
&&i\hbar\,\frac{\partial\varphi^{(B)}}{\partial t} = -\frac{\hbar^2}{2 m_B}
\nabla^2 \varphi^{(B)} + V_{trap}^{(B)} \, \varphi^{(B)}  \nonumber  \\
&&+\; g_B\, N_B\, |\varphi^{(B)}|^2 \, \varphi^{(B)}
+ g_{BF} \sum_{i=1}^{N_F} |\varphi_i^{(F)}|^2 \, \varphi^{(B)}
\nonumber  \\
&&i\hbar\,\frac{\partial\varphi_i^{(F)}}{\partial t} = -\frac{\hbar^2}{2 m_F}
\nabla^2 \varphi_i^{(F)} + V_{trap}^{(F)} \, \varphi_i^{(F)}
\nonumber  \\
&&+\; g_{BF}\, N_B\, |\varphi^{(B)}|^2 \, \varphi_i^{(F)}   \;.
\;\;\;\;\; i = 1,2,...,N_F
\label{equ}
\end{eqnarray}

All of the above equations have simple interpretation. Removing the last
terms in these equations (i.e., neglecting the mean-field interaction
energy between Bose and Fermi components in comparison with other energies)
one recovers the Gross-Pitaevskii equation for degenerate Bose gas and
the set of Schr\"odinger equations describing the noninteracting Fermi
system. It is easy to notice that when the Bose and Fermi gases attract
each other strongly enough the mean-field energy connected with this
attraction can overcome the repulsive mean-field energy for bosons. This
can happen provided both phases have a significant overlap. It means that
the presence of a degenerate Fermi gas changes the character of the
interaction between bosonic atoms from repulsive to attractive. Therefore
the formation of bright solitons becomes possible.

Based on the above considerations one can easily write the
condition for the value of critical strength of attraction between
bosons and fermions. It is given by $g_B n_B = |g_{BF}^{cr}| n_F$,
where $n_B=N_B |\varphi^{(B)}|^2$ and $n_F=\sum_{i=1}^{N_F}
|\varphi_i^{(F)}|^2$ are the densities of both fractions,
normalized to the number of particles, taken at the center of the
trap. A rough estimation of $g_{BF}^{cr}$ can be found assuming
the densities of both components are calculated within the
Thomas-Fermi approximation and components do not interact. Then we
have
\begin{eqnarray}
|g_{BF}^{cr}| = C\, \frac{N_B^{2/5}}{N_F^{1/2}} \,g_B   \;,
\label{crit}
\end{eqnarray}
where $C=C_1\, (a_\perp^B/a_B)^{\,3/5}\,
(a_\perp^F/a_\perp^B)^{\,3}\, \lambda_B^{2/5}/\lambda_F^{1/2}$,
$C_1=3^{9/10}\, 5^{2/5}\, \pi /16 \approx 1.0$, $a_\perp$ is the
radial harmonic oscillator length, $\lambda=\omega_z /
\omega_\perp$ defines the aspect ratio of axially symmetric trap,
and $a_B$ is the s-wave scattering length for the pure Bose gas
related to the interaction strength through $g_B=4\pi\hbar^2
a_B/m_B$. The condition (\ref{crit}) has several implications.
Squeezing radially both Bose and Fermi components decreases the
value of critical $g_{BF}$. The same happens when the number of
fermions is getting bigger in comparison with the number of
bosons. For a particular trap numbers of atoms are limited by the
occurrence of a collapse \cite{collapse}. In the case of experiment
of Ref. \cite{collapse} it was found that the system was stable if
the number of atoms in both species ($^{87}$Rb and $^{40}$K in
$|2,2>$ and $|9/2,9/2>$ hyperfine states, respectively) were
smaller than $2 \times 10^4$. Taking parameters of that experiment
($\omega_\perp^B = 2\pi \times 215$ Hz and $\omega_z^B = 2\pi
\times 16.3$ Hz) and assuming the following numbers of atoms
$N_B=10^3$ and $N_F=10^4$, one gets the critical coupling
$|g_{BF}^{cr}| = 6.7\, g_B$ which equals the natural value of
$g_{BF}$ for $^{87}$Rb -$^{40}$K mixture in the double-polarized
state mentioned above. It is understood that the relation
(\ref{crit}) is only the necessary condition for creation of
bright solitons. Another important factor is the geometry of the
system.

\begin{figure}[thb]
\resizebox{2.9in}{2.3in}
{\includegraphics{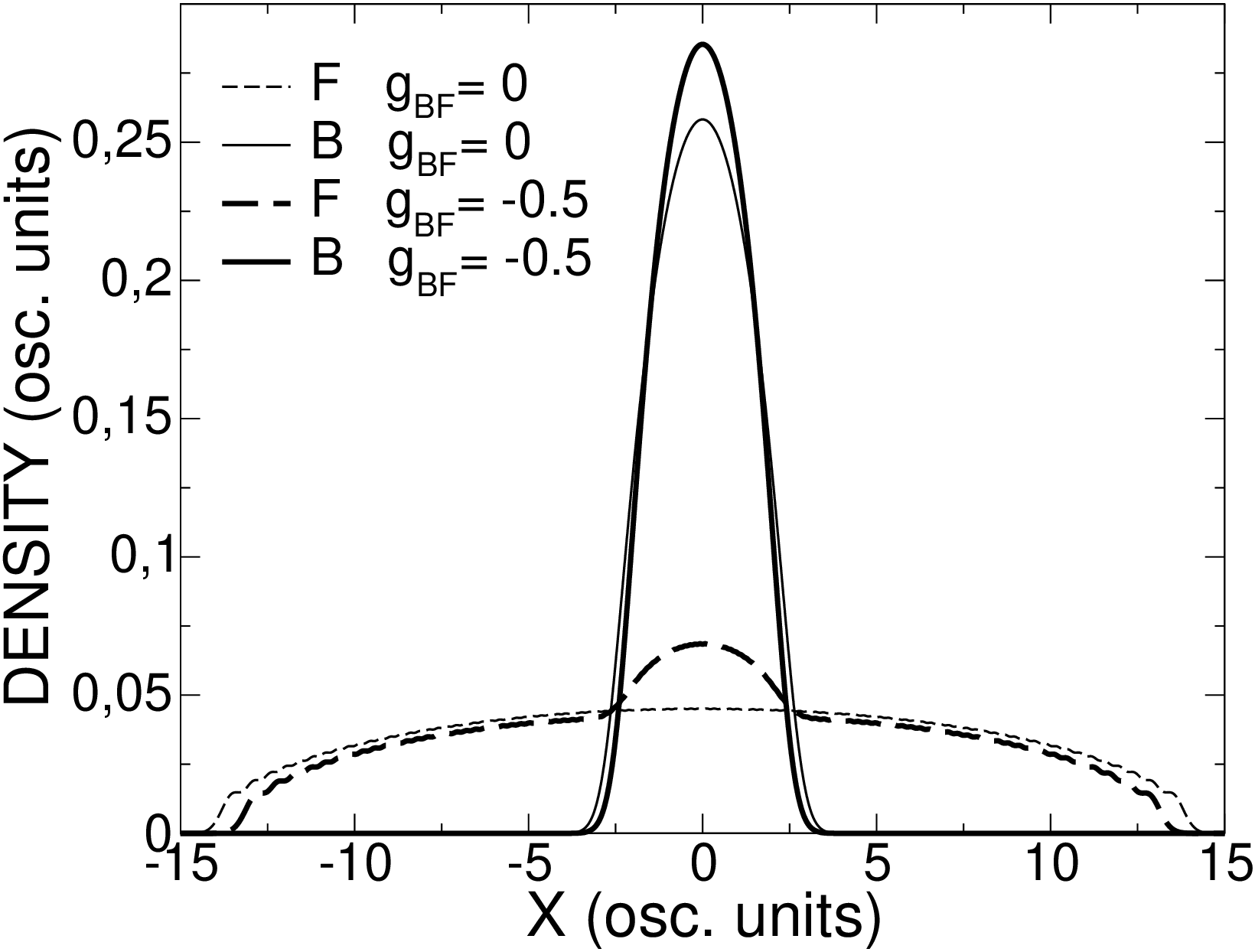}}
\caption{Density profiles (normalized to one) of one-dimensional
Bose-Fermi mixture consisting of $N_B = 1000$ bosons and $N_F = 100$
fermions for various strengths of attraction between components as
described in the label. The effective repulsion for bosons is governed
by $g_B=0.0163$. Solid lines correspond to the Bose fraction whereas the
dashed ones indicate fermions. By increasing the attraction between
components some fermions are pulled inside the Bose cloud. All quantities
are given in oscillatory units calculated based on fermionic component.}
\label{states}
\end{figure}
In the following we will concentrate on one-dimensional geometry, 
which is most favorable for the appearance of solitons. In \mbox{Fig.
\ref{states}} we plot the density profiles of Fermi (dashed lines)
and Bose (solid lines) components of one-dimensional mixture in
its ground state. The gases are confined in traps with frequencies
$2\pi \times 30\,$Hz (for fermions) and $2\pi \times 20\,$Hz (for
bosons). To get these curves we solve numerically the set of Eqs.
(\ref{equ}) by evolving adiabatically the coupling constant
$g_{BF}$ from zero to the given value. Even though the number of
bosons is $10$ times larger than the number of fermions, due to
the Pauli exclusion principle the size of the fermionic cloud is
much bigger. We see that after turning on the attractive forces
between components some number of fermions is drawn inside the
bosonic cloud. When the attraction is increasing further the
fermionic cloud is clearly divided into two distinguishable parts.
One of them is a broad background gas whereas the second is the
narrow density peak hidden within the bosonic peak. Both peaks get
higher and narrower when the attraction is getting stronger. This
is a sign of effective attraction.

\begin{figure}[thb]
\resizebox{2.5in}{3.in}
{\includegraphics{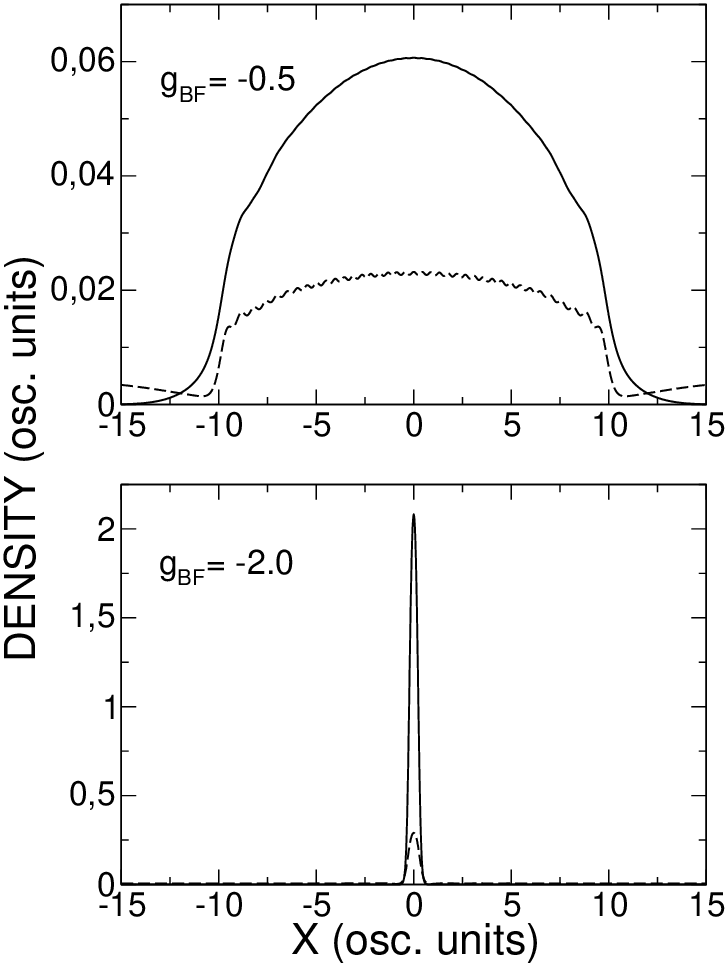}}
\caption{Illustration of the solitonic character of the ground state of
Bose-Fermi mixture. Both frames show the densities $34$ ms after
switching off the trapping potential. Only when the attraction between
the species is stronger than the critical one (the lower frame) the system
forms double-peak structure which does not spread in time.}
\label{open}
\end{figure}
It turns out that for strong enough attraction between Bose and
Fermi components the central peaks (bosonic and fermionic ones) in
Fig. \ref{states} form, in fact, the soliton. After switching off
the trapping potential the fermionic background is lost but the
bosonic peak persists without changing its shape and confines
fermionic density whose profile is also preserved. It is
illustrated in Fig. \ref{open}, where fermionic and bosonic
densities are plotted some time after the opening the trap. Upper
and lower frames differ by the value of $g_{BF}$. In the case of
the upper frame the strength of attraction is weaker than the
critical one and no solitonic behavior is observed - the bosonic
and fermionic clouds spread out. For the lower frame the
attraction is strong enough and the formation of a bright soliton
is observed. Such a structure can be forced to move by imposing a
momentum on it (realized experimentally by applying the Bragg
scattering technique) and its shape again does not change. The
critical value of the coupling constant is in our one-dimensional
model approximately equal to $g_{BF}^{cr} \approx -1.0$ osc. units
and can be compared to the value obtained based on the
one-dimensional counterpart of condition (\ref{crit}). Taking from
Fig. \ref{states} the ratio $n_B(0)/n_F(0)=57.5$ one gets
$g_{BF}^{cr} = -0.94$ osc. units what remains in agreement with
numerical estimation.

\begin{figure}[htb]
\resizebox{2.5in}{3.in}
{\includegraphics{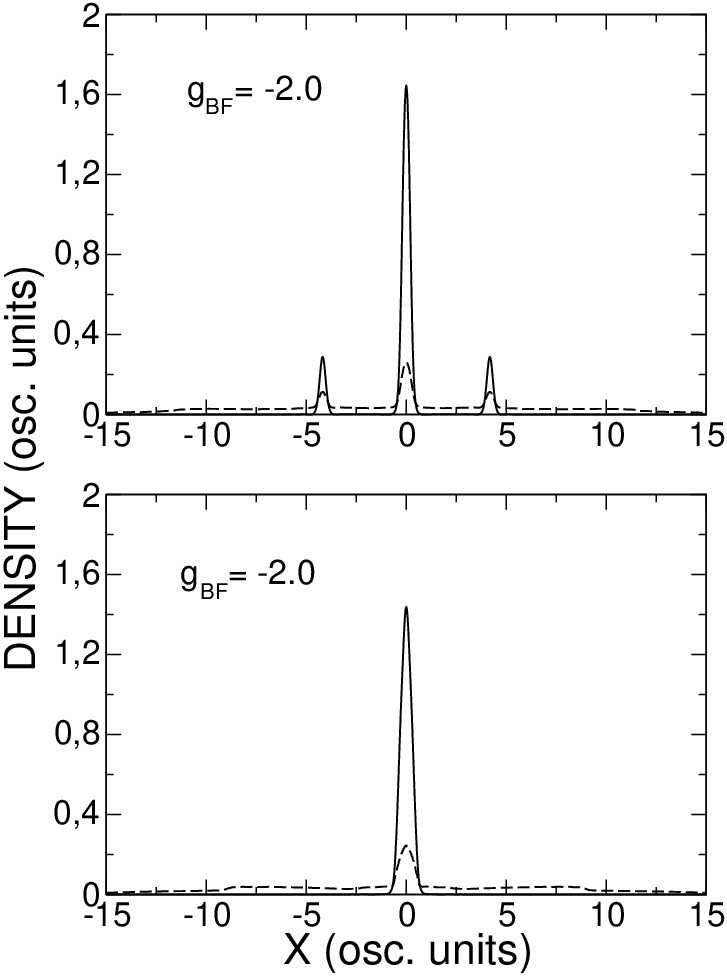}}
\caption{Density profiles of one-dimensional Bose-Fermi mixture after
switching the attraction between Bose and Fermi components from
$g_{BF} = -0.5$ to  $g_{BF} = -2.0$ oscillatory units. The interaction 
strength is changed linearly during $8.5$ ms (the upper frame) and
$17$ ms (the lower frame). The snapshots are taken at $20$ ms. More 
adiabatic change of the strength of attraction results in lower number 
of solitons. }
\label{turnon}
\end{figure}
\begin{figure}[thb]
\resizebox{2.5in}{3.in}
{\includegraphics{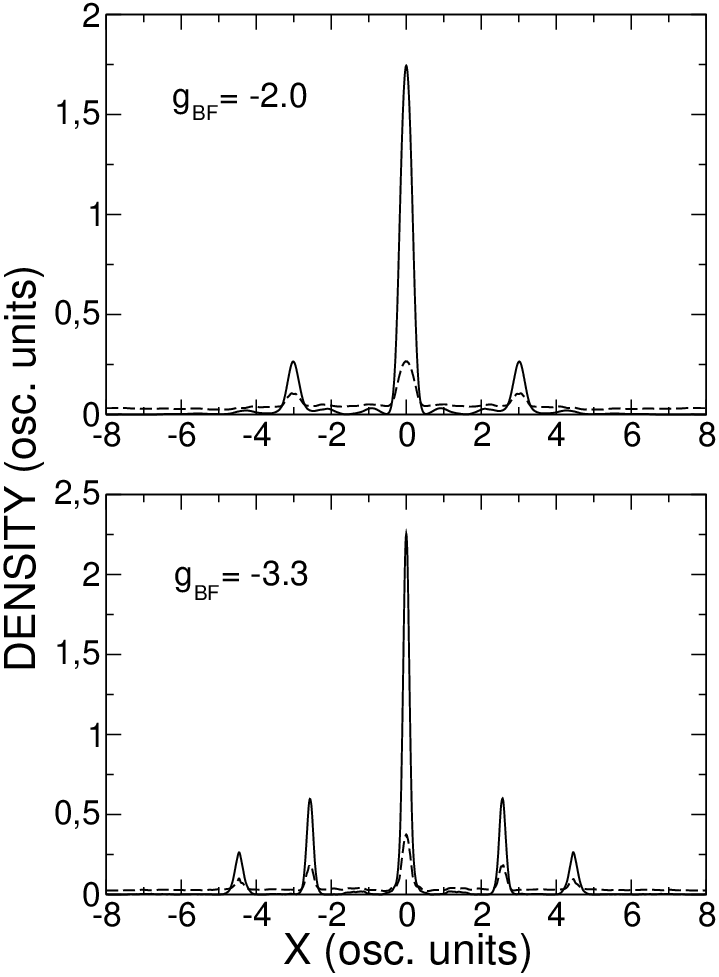}}
\caption{Density profiles of one-dimensional Bose-Fermi mixture $10.5$ ms
after changing the strength of attraction between fractions from the
initial value $g_{BF} = -0.8$ osc. units to the final one indicated
by the label. Here, the strength of the interaction is increased
instantaneously. The bosonic cloud breaks into several bright solitons,
each soliton contains a piece of fermionic cloud. The bigger jump in the
coupling $g_{BF}$ results in a bigger number of solitons.}
\label{solitons}
\end{figure}

In Figs. \ref{turnon} and \ref{solitons} we show the densities of
Bose and Fermi components after the strength of attraction between
bosons and fermions has been increased by using the Feshbach resonance 
technique. The other way of changing the interaction strength could be 
the firm radial squeezing of both samples. The basic observation is that 
after some time after switching the mutual interaction the bosonic cloud
breaks into several peaks provided the final coupling (attraction)
between both components is strong enough. Stronger attraction
results in bigger number of peaks. Each bosonic peak (marked by
the solid line) contains the fermionic density (marked by the
dashed line). Such double-peak structures oscillate almost without
changing their shapes. Switching off the trap shows that the
observed structures are indeed solitons. In the case of Fig.
\ref{solitons} the strength of the coupling between Bose and Fermi
gases is changed instantaneously. The contrast of the peaks depends
on the initial width of the bosonic cloud and it gets higher for
bigger number of bosons or smaller bosonic trap frequency. The latter
can be realized only in the optical trap. On the other hand, in Fig.
\ref{turnon} we show the response of the system when the mutual
attraction is increased within the finite time which is of the
order of trap period. Slower switching of the interaction leads to
a smaller number of solitons.

Based on numerical results, we propose two schemes for generating bright
solitons in degenerate Bose-Fermi mixtures. First of all, the system
has to be pushed in the range of strong enough attraction between fermions
and bosons. This can be achieved by producing a mixture in an appropriate
doubly-polarized state (for details see \cite{Simoni}) and by using the
Feshbach resonance technique. Another way is to follow the idea reported in
Ref. \cite{Olshanii} according to which squeezing radially both samples
firmly enough imposes a one-dimensional geometry on the system with
one-dimensional interaction strength given by $g_{3D}/\pi a_\perp^2$
($a_\perp$ is the radial harmonic oscillator length). Increasing radial
confinement leads then to higher effective one-dimensional interaction
strength. Therefore, one way to generate solitons in Bose-Fermi mixture
would follow the scenario: (a) the ground state of the system is formed in
elongated trap at the natural value of $g_{BF}$ (b) the strength of attraction
$g_{BF}$ is then increased. Another way could be performing the evaporation
already under favorable conditions, i.e., at the presence of the appropriate
magnetic field or strong enough one-dimensional geometry. However, in this
case only one, placed at the center of the trap double-peak soliton is formed.

In conclusion, we have shown that bright solitons can be generated
in a Bose-Fermi mixture as a result of a competition between two
interparticle interactions: boson-boson collisions which are
effectively repulsive and boson-fermion collisions which are
attractive. Assuming the strength of attraction is large enough
both kinds of atoms start to mediate in the other species
interaction introducing the system into a new regime where locally
Bose and Fermi gases become gases of effectively attractive atoms.
Therefore it becomes possible to generate bright solitons in the
system under such conditions. Depending on how fast the change of
the attraction strength is performed the system responses forming
the train of solitons (fast change) or the single soliton
(adiabatic change). Each soliton is, in fact, the double-peak
structure with the fermionic cloud hidden within the bosonic one.

\acknowledgments We are grateful to C. Ospelkaus and K. Sengstock
for stimulating discussions. M.B. and M.G. acknowledge support by
the Polish KBN Grant No. 2 P03B 052 24. T.K. and K.R. were
supported by the Polish Ministry of Scientific Research Grant
Quantum Information and Quantum Engineering No.
PBZ-MIN-008/P03/2003. S.O.-S. and K.B. acknowledge support by the
Deutsche Forschungsgemeinschaft in Schwerpunktprogramm SPP1116.

\bibliographystyle{apsrev}

\end{document}